# K-CORONA RECORDING IN THE RANGE < 1.4 R☉

Kim I.S.[1], Bugaenko O.I.[1], Lisin D.V.[2], Nasonova L.P.[1]
1- Lomonosov Moscow State University, Sternberg Astronomical Institute
1 -  IZMIRAN
*kim@sai.msu.ru*

*Two approaches are suggested for recording the continuum corona in the range < 1.4 R☉. They are different from the classical coronagraphic ones. Current state in the thin film technology allows discussing a new generation coronagraph with a variable transmission of an entrance aperture. The estimated coronagraphic factor is 2 orders of magnitude higher compared to a Lyot-type coronagraph. Another approach is based on the use of total solar eclipses at near-Mercury orbits. The instrumental background is decreased at least 3 orders of magnitude. That allows using a more simplified optical sketch.*

## 1. Introduction

The range < 1.4 R☉ (distances are counted from the solar disc center) is widely presented by ground and space data of the neutral and ion (E-corona) components in optical, UV, SXR, and HXR spectral intervals. The direct recording of the electron component (K-corona emitting in optical continuum) in the same range is carried out only during total solar eclipses (TSE). K-corona emission is linearly polarized and explained by scattering by free resting electrons. Indirect images are derived from measurements of the polarized brightness ($p$B, K-coronometer of Mauna Loa observatory – MK4 MLSO, USA).

The sky brightness and the instrumental background caused mainly by diffraction of the bright solar disk at the entrance aperture are the main factors complicating the recording of K-corona. The use of classical coronagraphs reduces the instrumental background to $\approx 10^{-5}$ B☉, where B☉ is the brightness of the solar disc center. The optical sketch of the Lyot coronagraph [Lyot 1931] has the internal occulting elements: the primary optics, the field lens, masking in the primary focal plane and masking in the plane of the image of the entrance pupil, and the relay optics.

Space coronagraphs use the external and internal masking and selected spectral intervals and start to record in the range > 1.4 R☉. Future space missions Proba 3 and Solar Orbiter plan to record in the range < 1.4 R☉. Below, we discuss two approaches for the direct K-corona recording, which are different from the Lyot method.

## 2. Apodizing with a mask in the plane of an entrance aperture

Diffraction at the entrance aperture of the primary optics is the main factor complicating the successful recording of faint objects near the bright ones (in our case – the Sun). As a result, the significant instrumental background appears and exceeds the K-corona brightness by 2-3 orders of magnitude.

The diffraction pattern in the focal plane is a result of the discontinuity of the transmission function (or its derivatives) at the entrance aperture. For the round aperture that is the discontinuity of the transmission function: $G(\rho)=1$ in the range $\rho<1$ and $G(\rho)=0$ for $\rho>1$, where $\rho$ is the distance from the center of the round aperture. The use of the mask with a variable transmission $[G(\rho)=1-\rho^2]$ (the discontinuity for the first derivative) significantly reduces the instrumental background [Kim et al. 2013]. According to preliminary estimations for solar upper atmosphere observations (R☉ = 960″, $\lambda$ = 600 nm, diameter of the entrance aperture D = 200 mm), the expected theoretical coronagraphic factor exceeds the Lyot one by 1-2 orders of magnitude at the chromosphere level ($h$ = 4″), and by 4 orders of magnitude at prominence heights ($h$ = 40″). Successes of current technology allow us to discuss the development of a "coronagraph", whose sketch consists of the primary optics with variable transmission and the recording assembly. The transmission of such a coronagraph is only 3 times less as compared with the Lyot type coronagraph.

### 3. Ground-based and space-borne total solar eclipses

The second approach is based on the use of total solar eclipses (TSE). During the totality a naked eye or a simple camera operate as an ideal external occulting coronagraph. The external occulting disk, the Moon, is at "infinity", allowing to approach the most inner part of the corona. K-corona is easily observed by the naked eye up to 10 R☉, as the sky brightness and the instrumental background are minimal. Let us estimate the instrumental background for the Earth-based TSE. The expression by Lensky [Lensky 1981] for the single external occulting disk is used to estimate the brightness at axial point of the primary optics.

$$E_0 = \left( \pi^2 i_{sun} \left[ 1 - \left( \frac{i_{sun}}{i'} \right)^2 \right] \right)^{-1} \times \frac{\lambda}{\rho},$$

where $E_0$ (the luminosity) is given in the units of the brightness of the solar disc, $i_{sun}$ is the angular radius of the solar disc, $i'$ is the angular radius of the external occulting disc (the Moon), $\lambda$ is the wavelength, $\rho$ is the radius of the external occulting disk (the Moon). Below the term "brightness" is used instead the luminosity as the expression is valid for the brightness in the focal plane. Hereinafter brightness and intensity are given in the units of the solar disc center ones. The calculated dependence of $E_0$ on eclipse magnitude $m$ ($i'/i_{sun}$) is shown in Figure 1 (left) for ground TSE: $\lambda$ = 550 nm (the grey curve), and $\lambda$ = 660 nm (the black curve). For typical $m$ = 1.02-1.08, $E_0 = 4\times10^{-11} - 2\times10^{-10}$. In practice, $E_0$ is degraded by 1-3 orders of magnitude by terrestrial atmosphere (air pollution, aerosols), the altitude of an observational site, etc. Taking into account the above-mentioned, the reasonability of the use of TSE in space is evident, in particular, at near planet orbits.

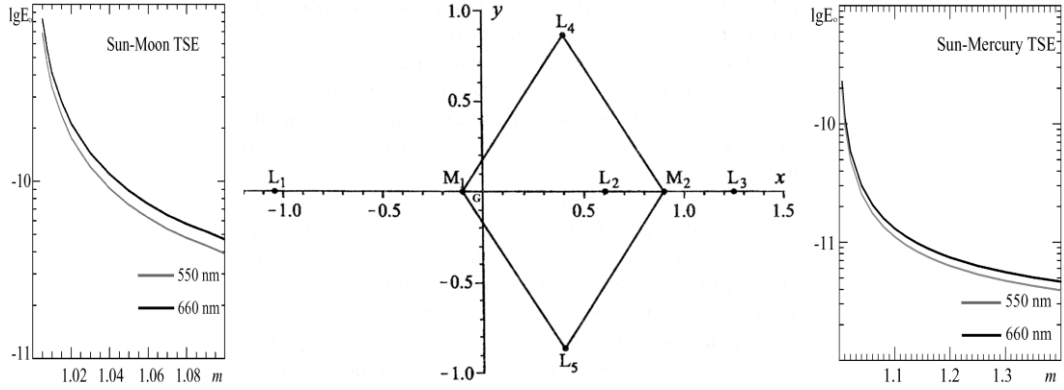

Figure 1.

***Let us consider conditions for recording the K-corona for terms orientation Sun-Planet-Spacecraft for the Lagrange points for Mercury.*** The motion of the spacecraft in the gravitational field of the Sun and any planet is a limited three-body problem, if we ignore the gravity attraction of other planets in the solar system [Lukianov and Shirmin 2009]. In this problem, there are exact solutions, found almost simultaneously by Euler in 1767 (the so-called straight, or collinear, or Euler libration points) and Lagrange in 1772 (the so-called triangular libration points). In a barycentric coordinate system Gxyz, rotating at a constant angular velocity *n*, equal to the angular velocity of the main bodies relative to each other, the main masses M1 and M2 (Sun and the planet) are located on the Gx axis (Figure 1, center). The Gz axis is directed along the vector of the angular velocity *n*. In this case the solutions are stationary points (i.e., points with the constant coordinates) called Lagrange libration points. Collinear libration points are located approximately along the axis Gx: L1 …. M1 .... G .... L2 .... M2 .... L3. M2 planet is located between two libration points L2 and L3, and the triangular libration points form in the Gxy plane two equilateral triangles, symmetric about the Gx axis with a common side on M1 M2 segment. The numbering of the linear libration points is not universally accepted. For our task, the use of TSE in space, the libration point L3 is suitable. For an observer at the point L3, the planet M2 will permanently occult the Sun (M1). The angular sizes of the Sun and Mercury are comparable.

For example, Figure 1 (right) shows the calculated instrumental background in the focal plane of the primary optics of a space telescope located near the collinear libration point L3 for Mercury. The Mercury radius of 2440 km and the distance of $5.79 \times 10^7$ km were used. The grey and black curves correspond to observations in the green and red spectral intervals. The calculated instrumental background in the range 1.05-1.4 R varies in the range $2\times10^{-11} - 4\times10^{-12}$.

It is known that the instrumental background is reduced by more than three orders of magnitude when $m$ ($i'/i_{sun}$) > 0.9 [Nikolsky 1972].

*Conclusions*

Preliminary estimations show opportunity for the K-corona recording in the range < 1.4 R☉ using two approaches different from the classical Lyot method.
- ✓ The use of a primary optics with variable transmission. The direct K-corona recording becomes possible with a telescope consisting of a primary optics and a recording assembly.
- ✓ The use of TSE at near planet orbits. There should be the only masking in the primary focal plane when $m$ ($i'/i_{sun}$) > 0.9, the optical sketch of a telescope is simplified.

The study was partially supported by RFBR (research project # 14-02-01225).

**References**

1. Lensky A.V. 1981. Azh. **58**. 648.
2. Lukianov L.G., Shirmin G.I. 2009. Lekzii po nebesnoi mekhanike. Uchebnoe posobie dlya vysshikh uchebnykh zavedenii. Almaty. Kazakhstan. 276 P (in Russian).
3. Nikolsky G.M. 1972. Byulletin Solnechnye Dannye Akademii Nauk USSR, No. 12. 96 (in Russian).
4. I.S. Kim, I.V. Alexeeva, O.I. Bugaenko, V.V. Popov, E.Z. Suyunova. 2013, Solar Phys. **288**, 2, 651-661, 2013, doi 10.1007/s11207-013-0419-0.
5. B. Lyot. C.R. Acad. Sci.**193**. 1169